\documentstyle[12pt]{article}
\begin{document}
\begin{center}
\begin{large}
{\bf A Variational Principle for the Asymptotic Speed of
\\ Fronts
of the Density Dependent Diffusion-Reaction Equation}
\end{large}
\end{center}
\vspace{.1cm}
\begin{center}
R.\ D.\ Benguria and M.\ C.\ Depassier\\ Facultad de F\'\i
sica\\ P. Universidad Cat\'olica de Chile\\ Casilla 306,
Santiago 22, Chile
\end{center}

\date\today
\begin{abstract}
We show that the minimal speed for the existence of monotonic
fronts of the  equation $u_t = (u^m)_{xx} + f(u)$ with
$f(0)=f(1)=0$, $m>1$ and $f>0$ in $(0,1)$, derives from a
variational principle.    The variational principle allows to
calculate, in principle, the exact speed for general  $f$. The
case $m=1$ when $f'(0) = 0$ is included as an extension of the
results.

\end{abstract}

\newpage

 Several problems arising in population growth
\cite{Newman80,Aronson}, combustion theory
\cite{Vulis,Clavin94}, chemical kinetics \cite{SS92}, and others
\cite{Sagan}, lead to an equation of the form
$$
{\partial\rho\over \partial t} + \vec\nabla \cdot \vec \jmath =
F(\rho),
$$
where the source term $F(\rho)$ represents net growth
and saturation processes.  The flux $\vec\jmath $ is given by
Fick's law
$$
\vec \jmath = - D(\rho ) \vec\nabla\rho,
$$
where the diffusion coefficient $D(\rho )$ may depend on the
density or in simple cases be taken as a constant. In one
dimension this leads to the equation
$$ {\partial\rho\over
\partial t} =  {\partial\over \partial x} \left( D(\rho)
{\partial\rho\over\partial x} \right) + F(\rho). \eqno(1a)
 $$
 In
what follows we shall assume that
$$
 F(\rho) > 0 \quad {\mbox{\rm in}} \quad (0,1),
\quad{\mbox{\rm and}} \quad F(0) = F(1) =0,\eqno(1b)
$$
restrictions which are satisfied by several models. When the
diffusion coefficient is constant and the additional requirement
$F'(0)>0$ is satisfied, the asymptotic speed of propagation of
localized small perturbations to the unstable state $u=0$ is
bounded below and in some cases coincides \cite{AW78} with the
value $c_L = 2\sqrt{F'(0)}$ which is obtained from
considerations on the linearized equation \cite{KPP37}.
However, when either $F'(0)=0$ or $D(\rho )$ is not a constant,
no hint for the speed of propagation of disturbances can be
obtained from linear theory alone. A common choice for the
diffusion coefficient is a power law, case with which we shall
be concerned here. Therefore the equation that we study is
$$
{\partial\rho\over \partial t} =   \left( \rho^m  \right)_{xx} +
F(\rho) \eqno(2a)
$$
 with
 $$ F(0) = F(1) = 0, \quad {\mbox{\rm
and}}\quad  F>0 \quad {\mbox{\rm in}} \quad (0,1). \eqno(2b)
$$
Aronson and Weinberger \cite{AW78,Aronson} have shown that the
asymptotic speed of propagation of disturbances from rest is the
minimal speed $c^*(m)$ for which there exist monotonic
travelling fronts $\rho(x,t) = q(x - ct)$ joining $q=1$ to
$q=0$. The equation satisfied by the travelling fronts is
$$
(q^m)_{zz} + c q_z + F(q) = 0 \eqno(3a)
$$
 with
$$
q(-\infty)=1,
\quad q>0, \quad q' <0 \quad {\mbox{\rm in}}\quad
(-\infty,\omega), \quad q(\omega ) = 0 \eqno(3b)
$$
where $z= x
- ct$. The wave of minimal speed is sharp, that is, $\omega <
\infty$ when $m>1$ \cite{Aronson}.

An explicit solution is known \cite{Newman80,Aronson} for the
case $F(q) = q (1 -q)$ and $m =2$, the waveform is given by
$$
q(z) = \left[ 1 -{1\over2} {\rm e}^{z/2} \right]_+
$$
 and it
travels with speed $c^*(2) = 1$ (here $[x]_+ \equiv \max (x,0)$).
 Recently the derivative $dc/dm$
at $m=2$ has been calculated by two different methods. Its value
is -7/24 \cite{AV94,Chen94}. Other exact solutions for different
choices for $m$ and $F$ have been given in \cite{HMO92}.

The purpose of this work is to give a variational
characterization of the minimal speed $c^*(m)$ for Eq.(3) when
$m>1$, and as a byproduct for the case $m=1$ when $F'(0) =0$,
both, cases for which no information is obtained from linear
theory. The case $m=1$ with $F'(0)>0$ has been studied
elsewhere \cite{BDpreprint}. Lower bounds have been obtained on
the minimal speed $c^*(m)$ \cite{BD94}; the present results allow
its exact calculation for arbitrary $f$.

 Since the selected speed corresponds to that of a decreasing
monotonic front, we may consider the dependence of its
derivative $dq/dz$ on $q$. Calling $p(q) = - q^{m-1}dq/dz$,
where the minus sign is included so that $p$ is positive, we
find that the monotonic fronts are  solutions of
$$
 p {dp\over
dq} - {c^*\over m} p + {1\over m} q^{m-1} F(q) = 0 \eqno(4a)
$$
with
$$ p(0) = p(1) = 0, \quad p>0 \quad {\mbox{\rm in}}\quad (0,1).
\eqno(4b)
$$
 Although the wave of minimal speed is sharp and
therefore $q'(0) < 0$, by its definition $p(0) = 0$ is true. We
now show that the minimal speed $c^*(m)$ follows from a
variational principle whose Euler equation is  Eq.(4a).

Let $g$ be a positive function such that  $h = - g' > 0$.
Multiplying Eq.(4a) by $g/p$ and integrating we obtain after
integration by parts,
$$
 {c\over m} = {\int_0^1 \left[{1\over m}
q^{m-1} F(q) {g(q)\over p(q)} + h(q) p(q) \right]\, dq \over
\int_0^1 g(q)\, dq }.
\eqno(5)
$$
By Schwarz's inequality, since, $q$, $F$, $g$ and $h$ are
positive we know
$$
 {1\over m} q^{m-1} F {g\over p} + h p \ge 2
\sqrt{{1\over m} q^{m-1} F g h}
\eqno(6)
 $$
and therefore, replacing in Eq.(5) we have
$$
 c \ge 2 {\int_0^1
\sqrt{m q^{m-1} F g h} dq \over \int_0^1 g dq}. \eqno(7)
$$
 This
bound has been already given by us \cite{BD94}. We now show that
it is always possible to find a $g(q)$ such that the equality in
Eq.(6) and therefore also in Eq.(7) holds. We do so by explicit
construction of such a function $g$. The equality in Eq.(6)
holds if
$$
 {1\over m} q^{m-1} F {g\over p} = h p \eqno(8)
$$
Let $v(q)$ be the positive solution of
$$
 {v'\over v} = {c\over
mp} \eqno(9a)
$$
 and choose
$$
 g = {1\over v'}. \eqno(9b)
$$
 We
have then
$$
 {v''\over v} = {(v')^2\over v^2} - {c\over mp^2} p'
= {c\over m^2 p^3} q^{m-1} F(q)
$$
 where we have used Eq.(9a) to
eliminate $v'$ and Eq.(4a) to eliminate $p'$. Therefore,
$$
 h =
- g' = {v'' \over (v')^2} = {1\over m p^2} q^{m-1} F g > 0
\eqno(9c)
$$
 where we have made use of Eqs.(9a) and (9b).  With
this expression for $h$, we can see that Eq.(8) is satisfied. In
addition we must check that $g$ as we have defined it is such
that its integral exists. In fact as it exists and moreover one
can always normalize $g$ so that $g(0)=1$ and $g(1) = 0$. From
the definition of $g$ we obtain
$$
g(q) = {m p(q)\over c} \exp\left[-\int_{q_0}^q {c\over m p}\, dq'\right]
 $$
 where  $0< q_0 < 1$.  Since $p(1) = 0$ and $p$
is positive between 0 and 1 it follows that $g(1) = 0$. At zero
no divergence occurs, as we now show. Call $\hat c = c/m$ and
$f(q) = q^{m-1} F(q)/m$. Then Eq.(4a) reads
$$
 p p' - \hat c p +
f =0 \eqno(10a)
$$
 with
 $$
f(0) = f(1) = 0  \quad {\mbox{\rm
and}} \quad f'(0) =0. \eqno(10b)
$$
 For this case Aronson and
Weinberger \cite{AW78} have shown that $p(q)$ approaches the
fixed point $q=0$ as $ p = \hat c q =  c q/m$. Then, near 0,
$v'/v \approx 1/q$ or $v \approx q$ and from its definition
$g(0) =1$. Then the integral of $g$ exists. We have shown then
$$
 c^*(m) = \max 2\, {\int_0^1 \sqrt{m q^{m-1} F g h}\, dq \over
\int_0^1 g \,dq}. \eqno(11)
$$
 where the maximum is taken over all
functions $g$ such that
$$
 g(0) = 1, \quad g(1) = 0\quad
{\mbox{\rm and}} \quad h = - g' > 0.
$$
 It is perhaps of some
interest to verify explicitly that the Euler equation for the
maximizing $g$ is indeed Eq.(4a). Let us study the maximization
of  the functional
$$
 J_m(g) = 2 \int_0^1 \sqrt{m q^{m-1} F g h}\,dq
$$
 where $ h = - g' >0$ subject to
$$
\int_0^1 g(q)\, dq = 1.
$$
 The Euler equation for this problem is
$$
\lambda + \sqrt{{m q^{m-1} F h\over g}} +
 {d\over dq}\left( \sqrt{{ m q^{m-1} F g\over h}} \right) = 0
$$
 where $\lambda$ is the Lagrange multiplier.  Using the
expression given in Eq.(9c) for $h$ we see that this is exactly
Eq.(4a) with the Lagrange multiplier $\lambda = -c$.

As an application we shall consider the case $F(q) = q(1-q)$ and
$m=2$ for which the exact solution is known. Take as the trial
function $g(q) = (1 -q)^2$. Then we obtain
$$
 c \ge 4 {\int_0^1
q (1-q)^2 dq \over \int_0^1 (1-q)^2 dq} = 1
$$
the exact value,
which shows that this is the function $g$ for which the maximum
is attained. In addition, due to the existence of the
variational principle we may use the Feynman-Hellman formula to
calculate the dependence of $c(m)$ on parameters of $F$. We
illustrate this by applying it to the calculation of $dc/dm$ at
$m=2$.  Taking the derivative of Eq.(10) with respect to $m$ we
obtain
$$
 {dc \over  dm} =  {1 \over \int_0^1 g dq} \int_0^1
{ghF\over \sqrt{mF q^{m-1} g h}} [q^{m-1} (1 + m \log q) ] dq.
$$
Evaluating at $m=2$, with $g(q) = (1 -q)^2$ we obtain $$ {dc
\over dm}(2) = 3\int_0^1 q(1-q)^2 (1 + 2 \log q) dq = -{7\over
24} $$ the value previously obtained by other methods.

A fast estimation of the speed for other values of $m$ can be
obtained with simple trial functions. In Fig. 1 we show lower
bounds for $F = q(1-q)$ using as trial functions $g_1 = (1 -q)^2$
and $g_2 = (1-q)$. With the first trial function we have the
exact value at $m=2$. The dotted line is the line of slope -7/24
that coincides with the tangent at $m=2$. For larger $m$ a
better estimate is obtained using $g_2$. The dashed line is the
curve $\sqrt{2/m}$ which has been suggested by Newman
\cite{Newman80} as the best fit to his numerical results. With
better choice of trial functions the exact value can be
approached arbitrarily close.

Finally we observe that the case $m=1$ when $F'(0) = 0$ follows
directly here. Repeating the procedure starting now from
equation (10), one obtains,
$$
 c = \max 2\, {\int_0^1 \sqrt{ F g h}\, dq \over \int_0^1 g\, dq}
 $$
 where the maximum is taken over all
functions $g$ such that $$ g(0) = 1, \quad g(1) = 0\quad
{\mbox{\rm and}} \quad h = - g' > 0.  $$ To show this we have
used $v'/v = c/p$ and $g = 1/v'$ and the asymptotic behavior
described above.

\section{Acknowledgments}

We thank Prof. Dirk Meink\"ohn for giving us several useful
references.  This work was partially supported by Fondecyt
project 193-0559.

\end{document}